\documentclass[conference]{IEEEtran}

\usepackage{hyperref}
\usepackage[pdftex]{graphicx}
\usepackage{amsmath,amssymb}
\usepackage{xcolor}
\usepackage{colortbl}
\usepackage{eiffel}
\usepackage{xspace}
\usepackage[capitalize]{cleveref}
  \crefname{section}{Sect.}{Sect.}
  \Crefname{section}{Section}{Sections}
  \crefname{figure}{Fig.}{Fig.}
  \Crefname{figure}{Figure}{Figures}
  \crefname{listing}{Listing}{Listings}
  \Crefname{listing}{Listing}{Listings}
  \crefname{table}{Tab.}{Tab.}
  \Crefname{table}{Table}{Table}
\usepackage{threeparttable}

\newif\ifextended\extendedtrue

\newcommand{\eb}{EiffelBase\xspace}
\newcommand{\ebmbc}{EiffelBase+\xspace}
\newcommand{\eif}[1]{\mbox{\lstinline|#1|}}

\lstdefinestyle{sharpc}{
  language=[Sharp]C,
  morekeywords=[2]{
    AvlTree,
    BinarySearchTree,
    CollectionBase,
    CommonBinaryTree,
    Deque,
    DoublyLinkedList,
    Heap,
    OrderedSet,
    PriorityQueue,
    SinglyLinkedList,
  },
  keywordstyle=[2]{\color[HTML]{3333FF}\itshape},
}
\newcommand{\csharp}[1]{\mbox{\lstinline[style=sharpc]|#1|}}

\begin{document}

\title{What Good Are Strong Specifications?}

\author{\IEEEauthorblockN{Nadia Polikarpova$^\ast$ $\quad$ Carlo A.\ Furia$^\ast$ $\quad$ Yu Pei$^\ast$ $\quad$ Yi Wei$^\ast$ $\quad$ Bertrand Meyer$^{\ast,\dag}$}
\IEEEauthorblockA{
\begin{tabular}{c@{$\ \ $}c}
$^\ast$Chair of Software Engineering, ETH Zurich, Switzerland  &  
   $^\dag$ITMO National Research University, St.~Petersburg, Russia \\[0.2mm]
\multicolumn{2}{c}{
firstname.lastname@inf.ethz.ch
}
\end{tabular}
}
}

\maketitle

\bstctlcite{IEEEexample:BSTcontrol}

\begin{abstract}
Experience with lightweight formal methods suggests that programmers are willing to write specification if it brings tangible benefits to their usual development activities.
This paper considers \emph{stronger} specifications and studies whether they can be deployed as an incremental practice that brings additional benefits without being unacceptably expensive.
We introduce a methodology that extends Design by Contract to write strong specifications of functional properties in the form of preconditions, postconditions, and invariants.
The methodology aims at being palatable to developers who are not fluent in formal techniques but are comfortable with writing simple specifications.
We evaluate the cost and the benefits of using strong specifications by applying the methodology to testing data structure implementations written in Eiffel and C\#.
In our extensive experiments, testing against strong specifications detects twice as many bugs as standard contracts, with a reasonable overhead in terms of annotation burden and run-time performance while testing.
In the wide spectrum of formal techniques for software quality, testing against strong specifications lies in a ``sweet spot'' with a favorable benefit to effort ratio.
\end{abstract}

\IEEEpeerreviewmaketitle

\section{Introduction} \label{sec:introduction} 

\noindent
Many years of progress in the theory and practice of formal methods notwithstanding, 
writing software specifications%
\footnote{In this paper, we target \emph{formal} specifications of \emph{functional} properties.}
still seems to be ``disliked by almost everyone''~\cite{Parnas-FOSE}. 
In many cases, this disliking is a consequence of a high cost/benefit ratio---perceived or real---of writing and maintaining accurate specifications on top of the code. 
After all, developers \emph{will} write specifications as long as they are simple, have a straightforward connection with the implementation, 
and help them write and debug code better and faster. 
One example is Design by Contract~\cite{Meyer97,Chalin06}
where simple executable specifications, 
written in the same syntax as programming language expressions, 
support design, incremental development, and testing and debugging.
Another one is test-driven development~\cite{TDD-book},
where rigorously defined test cases play the role of specifications in defining correct and incorrect behavior.
Experiences with these techniques show 
that providing lightweight specifications is an accepted practice 
when it brings tangible benefits and integrates well with the overall development process.

But what about \emph{strong} specifications, 
which attempt to capture the entire (functional) behavior of the software? 
Should we dismiss them on the grounds 
that the effort required to write them is not justified against the benefits they bring in the majority of mundane software projects? 
This paper studies the impact of deploying strong behavioral specifications, 
in the form of \emph{contracts} (pre- and postconditions and class invariants),
for detecting errors in software using automatic testing.

Using strong contracts involves costs and possible benefits.
Among the former we have the programming effort necessary to write such strong specifications 
and the runtime overhead of checking them during execution.
The benefits may include finding more errors, finding more subtle
errors, finding errors more quickly, and exposing errors in ways that
are easier to understand and correct.  Our \textbf{contributions}
address the cost factors---by measuring and trying to mitigate
them---and assess the benefits:
\begin{itemize}
\item \cref{sec:mbc} presents a methodology to write strong specifications%
  ---extending our previous work~\cite{PFM10-VSTTE10}---%
  that does not require fluency in formal techniques 
  because it is an extension of such traditional practices as Design by Contract. 
  This is instrumental in reducing the programming effort associated with strong specifications.

\item The methodology comes with tool support and specification libraries, so that strong specifications are usable with standard debugging and testing tools.

\item \cref{sec:experiment_setup,sec:results} describe an extensive empirical study 
  that evaluates the use of strong contracts for real software 
  and measures their costs and benefits in terms of defect detection.
\end{itemize}

The bulk of our empirical study targets \eb, 
a library of generic containers and data structures 
(such as lists, tables, and trees),
which has been in use in the Eiffel community for more than 20 years. 
The production version of \eb includes simple contracts, 
a form of partial specification, 
that are nonetheless quite effective at finding implementation bugs automatically using contract-based random testing~\cite{bertrand2009}, 
where executable contracts serve as oracles 
and enable a push-button testing process.
In the present paper, we augment the simple contracts that come with \eb using the methodology discussed in \cref{sec:spec_methodology}. 
The result is \ebmbc: a version of \eb with identical implementation but strong (mostly complete) specifications.

In an extensive set of experiments, we compare the effectiveness of random testing on \eb and \ebmbc, 
with the goal of assessing whether the additional effort invested into the strong contracts pays off in terms of quantity and complexity of the bugs found.
Our \textbf{experiments show} that these measures dramatically increase when deploying strong specifications:
random testing found twice as many bugs in \ebmbc, 
and the simple contracts of \eb would have uncovered none of the new bugs.
The overhead size of specifications, in contrast, remains moderate,
with the specification-to-code ratio going from 0.2 to 0.46.

Our approach to writing strong specifications that are effective for testing is not limited to Eiffel programs. 
In a companion set of experiments, we applied the same technique to writing strong specifications for the DSA C\# library~\cite{DSA} 
and tested the result using Pex~\cite{TillmannH08}; 
in this case too we discovered new bugs with reasonable additional effort.

\section{Strong Specifications: An Example} \label{sec:examples}

\noindent
The following example illustrates and justifies the use of strong specifications.
Consider the \eb class \eif{LINKED_LIST}---Eiffel's standard implementation of linked lists.
Like many containers in \eb, \eif{LINKED_LIST} includes an internal cursor to iterate over elements of the list.
The query\footnote{A \emph{query} is an attribute or a function~\cite{Meyer97}.} \eif{index} gives the cursor's position, 
which can be on any element of the list in positions 1 through \eif{count},
or take the special boundary values 0 (``\eif{before}'' the list) and \eif{count + 1} (``\eif{after}'' the list).
The attribute \eif{count} denotes the number of elements in the list.

\begin{table}
\begin{lstlisting}
merge_right (other: LINKED_LIST [G])
  require
    not after
    other /= Void
    other /= Current
  ensure
    count = old count + old other.count
    index = old index
  end
\end{lstlisting}
\caption{Standard specification of routine \eif{merge_right} in \eif{LINKED_LIST}.}
\label{tab:merge}
\end{table}

\cref{tab:merge} shows the \eb specification of \eif{LINKED_LIST}'s routine (method) \eif{merge_right}.
The routine inserts another list \eif{other} passed as argument into the current list (denoted \eif{Current} in Eiffel, corresponding to \lstinline[language=Java]|this| in Java and C\#) immediately after the cursor position.
For example, if \eif{Current} stores the sequence of elements {\footnotesize \textsf{b$\cdot$a$\cdot$r$\cdot$t}} 
with cursor positioned on the ``{\footnotesize \textsf{r}}'' (\eif{index = 3})
and \eif{other} stores {\footnotesize \textsf{o$\cdot$n$\cdot$e}},  
\eif{merge_right} changes \eif{Current} to {\footnotesize \textsf{b$\cdot$a$\cdot$r$\cdot$o$\cdot$n$\cdot$e$\cdot$t}}.
The precondition (\eif{require}) specifies that the routine cannot be called when the cursor is \eif{after}: there is no valid position to the right of it. 
It also demands that \eif{other} be non-\eif{Void} (\lstinline[language=Java]|null| in Java and C\#) and not aliased with the \eif{Current} list: otherwise, merging is not well defined.
The postcondition (\eif{ensure}) describes some expected effects of executing \eif{merge_right}: 
the \eif{Current} list will contain as many elements as it contained before the call to \eif{merge_right}
(denoted by \eif{old count}) 
plus the number of elements of the \eif{other} list; 
and the cursor's position \eif{index} will not change.

The contracts in \cref{tab:merge} are a good example of the kind of specification that Eiffel programmers normally write~\cite{Polikarpova2009}: 
it is correct and nontrivial, and it can help detect errors in the implementation, 
such as performing partial merges or incorrectly leaving the cursor at a different position.
Unfortunately the specification is also incomplete, 
because it does not precisely describe the expected state of the list after merging.
In fact, the current implementation of \eif{merge_right} contains an error that is undetectable against the specification of \cref{tab:merge}.
The error occurs in the special case of calling \eif{merge_right} with cursor \eif{before} the list (\eif{index = 0}): the implementation will insert \eif{other} at the second rather than at the first position.
For example, merging {\footnotesize \textsf{f$\cdot$o$\cdot$l$\cdot$d}} and {\footnotesize \textsf{u$\cdot$n}} when the cursor is \eif{before} 
yields {\footnotesize \textsf{f$\cdot$u$\cdot$n$\cdot$o$\cdot$l$\cdot$d}} instead of the correct {\footnotesize \textsf{u$\cdot$n$\cdot$f$\cdot$o$\cdot$l$\cdot$d}}.

\begin{table}
\begin{lstlisting}
merge_right (other: LINKED_LIST [G])
  require
    -- As in $\text{\cref{tab:merge}}$
  modify sequence
  ensure
    sequence = old (sequence.front (index) + 
                    other.sequence + sequence.tail (index + 1))
  end
\end{lstlisting}
\caption{Model-based specification of routine \eif{merge_right} in \eif{LINKED_LIST}.}
\label{tab:merge_mbc}
\end{table}

\cref{sec:mbc} presents a methodology to write, with moderate effort, strong specifications 
that extend and, whenever possible, complete this kind of partial specification.
\cref{tab:merge_mbc} shows the strong specification obtained by applying the methodology to \eif{merge_right}, 
the way it appears in \ebmbc.
As is common in most Eiffel projects, the programmer who wrote \eif{merge_right} did a good job with the precondition,
which is sufficiently detailed and need not be strengthened.
The postcondition, however, turns into a single assertion 
that defines the \eif{sequence} of elements stored in the list after calling \eif{merge_right} as the concatenation (operator \eif{+}) of three segments: 
\eif{Current}'s original sequence up until position \eif{index} (written \eif{sequence.front (index)}), 
followed by \eif{other}'s element sequence, 
followed by the original sequence from position \eif{index + 1} (written \eif{sequence.tail (index + 1)}).
This postcondition relies on an \emph{abstract model} of the linked list in the form of a mathematical sequence of elements, 
which was already implicitly present above, in the informal description of the semantics of \eif{merge_right}.
Models blend well with Eiffel's standard specification constructs to help formalize programmers' intuitive understanding of data structures semantics.
Using the strong postcondition in \cref{tab:merge_mbc}, 
completely automatic testing with the AutoTest tool~\cite{bertrand2009} detected the error 
that occurs in \eif{merge_right} when the cursor is \eif{before}.

The postcondition in \cref{tab:merge_mbc} describes how the sequence changes, but it does not say what does \emph{not change}.
Including the assertion \eif{index = old index} from the original postcondition is not sufficient, as it only mentions one piece of state that does not change.
Instead we include the assertion \eif{modify sequence}, which means that \eif{merge_right} may only modify the sequence of elements in the \eif{Current} list and \emph{nothing else}.
Together pre-, postcondition, and modify clause give a complete specification of \eif{merge_right} behavior, against which we can automatically test any implementation for correctness.

\section{How to Write Strong Specifications} \label{sec:mbc} \label{sec:spec_methodology}

\noindent
Writing good specification is hard; at least this is the common belief.
Experience with Design by Contract 
suggests that programmers can competently write simple specifications 
if they can be expressed using familiar syntax.
See for example the specification in \cref{tab:merge}, which refers to regular class queries such as \eif{count} and \eif{index}, also used in the implementation.

Without further guidance and language support, however, programmers tend to write only partial specifications, 
because expressing complex properties is cumbersome.
This section describes \emph{model-based contracts} (MBC): 
a methodology to write strong specifications that structures and extends traditional Design by Contract.
MBC includes simple guidelines to define the abstract model of a class (\cref{sec:abstr-class-models}), and to write pre- and postconditions of routines (\cref{sec:preconditions,sec:postconditions}) and other, more advanced, specification elements (\cref{sec:framing-properties,sec:class-invariants}).

The MBC approach supports writing strong specifications in a number of ways:
\emph{models} facilitate choosing the right level of abstraction and expressing complex behavioral properties concisely,
while the structured discipline for writing postconditions and invariants, 
together with the notion of \emph{completeness} (\cref{sec:framing-properties}),
provides precise guidelines as to which properties are worth documenting in a contract, and when a contract is strong enough.
While fostering rigor and accuracy in specifications, 
MBC is still palatable to practitioners because its notation is part of the programming language.
When developing specifications for testing, as opposed to formal verification, MBC can be exploited \emph{incrementally}:
developers may skip writing the most advanced specification elements (for example, complex class invariants) 
while still getting strong specifications that are useful to detect subtle errors.

The following subsections present MBC using examples from \eb.
The few additional constructs introduced by MBC are highlighted in a different color and underlined in the examples (e.g., \eif{modify}).
The current presentation of MBC derives from previous work of ours~\cite{PFM10-VSTTE10}, 
which focused on using strong specifications when designing new software.
In this paper we adapt the principles introduced in~\cite{PFM10-VSTTE10} 
to the goal of supplying \emph{existing software} with flexible strong specifications for runtime checking and automatic testing (see \cref{sec:mbc_at_runtime}).
We also extend the specification methodology with new construct that handle framing (\cref{sec:framing-properties}) and complex class invariants (\cref{sec:class-invariants}).

\subsection{Abstract Class Models} \label{sec:abstr-class-models}

\noindent
Writing strong specifications becomes simpler if we can readily express the \emph{abstract state space} of classes and how it changes.
Therefore, the first step in specifying a class with MBC is defining a \emph{model} for the class: a set of mathematical elements that capture the abstract state space.

\begin{table}
\ifextended
\begin{lstlisting}
class LINKED_LIST [G]

  model sequence, index

  sequence: MML_SEQUENCE [G]
    status specification
    -- Specification query: sequence of elements in the list.

  index: INTEGER
    -- Internal cursor position.

  off: BOOLEAN
      -- Is the cursor not on a list element?
    ensure
      Result = not sequence.domain.has (index)
    end

invariant
  -- Model constraint
  0 <= index and index <= sequence.count + 1
  -- Attribute definition
  count = sequence.count
  -- Linking invariant
  bag = sequence.to_bag
  -- Internal representation constraint
  not sequence.is_empty implies last_cell.item = sequence.last
end
\end{lstlisting}
\fi
\caption{Excerpt of \eif{LINKED_LIST}'s MBC specification in \ebmbc.}
\label{tab:llist_mbc}
\end{table}

Syntactically, the annotation \eif{model} (see \cref{tab:llist_mbc}) declares the abstract model of a class as a list of attributes or functions called \emph{model queries}; 
each element listed after \eif{model} is either a query of basic type (Boolean, integer, or object reference) already used in the implementation, 
or a \emph{specification query}, meaning a query introduced solely to define the model.
As part of our work on MBC, we developed the Mathematical Model Library (MML), 
a collection of immutable Eiffel classes that represent mathematical concepts useful for specification:
sets, bags, sequences, maps, and relations.
Specification queries make use of MML classes to represent complex components of class models.
For example, \eif{LINKED_LIST}'s model in \cref{tab:llist_mbc} has two components: a specification function \eif{sequence} with return type \eif{MML_SEQUENCE} that gives the abstract sequence of elements stored in the list, and the ordinary class attribute \eif{index} of integer type.

Class models should be expressive enough to formalize the class behavior as seen at the API level, 
without exposing implementation-specific details.
\ifextended
  For example, the same abstract model---a sequence of elements---is suitable for all three implementations of lists in \eb:
  singly-linked, doubly-linked and array-based,
  as the particular representation does not influence the functional properties of public routines.
\fi
In practice, it is usually easy to devise a model for a data structure using MML abstractions. 
Even for classes representing complex real-world concepts, such as an ATM or a flight scheduler, 
MML remains applicable if used incrementally to define partial yet useful behavioral properties.

\subsection{Preconditions} \label{sec:preconditions}

\noindent
The \emph{precondition} of a routine defines when a call to the routine is valid.
In practice preconditions appear to be the most widely and accurately used form of contract~\cite{Polikarpova2009}.
Therefore, MBC does not introduce special guidelines for writing preconditions.

\subsection{Postconditions} \label{sec:postconditions}

\noindent
The \emph{postcondition} of a routine~\eif{r} describes the intended effects of executing \eif{r} on the object state; it is a \emph{relation} between the state just before (denoted using the keyword \eif{old}) and the state just after executing \eif{r}.

MBC postconditions express the intended effect of executing a routine on the \emph{model}, that is in terms of the model queries.
Procedure \eif{merge_right} in \cref{tab:merge_mbc}, for example, declares its effect on the model query \eif{sequence} of the current object.
For \emph{functions}, the postcondition also mentions the returned object (and its model queries) using the keyword \eif{Result}.
For example, function \eif{off} in \cref{tab:llist_mbc} defines \eif{Result} in terms of \eif{sequence} and \eif{index}.

\subsection{Framing Specification} \label{sec:framing-properties}

\noindent
An accurate routine specification should limit the effects of the routine execution to a certain part of the program state. 
Such specification elements are called \emph{framing specifications}.

\ifextended
Eiffel offers no dedicated language support for writing framing specifications.
In principle this support is not strictly necessary, because one can express the unchanged elements in postconditions with expressions such as \eif{index = old index} in \cref{tab:merge}.
In practice, however, this is cumbersome because any given routine usually affects only a handful of program elements; hence explicitly specifying all that does not change is verbose and tedious.
In fact, Eiffel practitioners rarely write framing specifications in this form.
\fi

In MBC, the keyword \eif{modify} introduces a routine's framing specification: a list of all model queries whose value is allowed to change after executing the routine.
For example, routine \eif{merge_right} in \cref{tab:merge_mbc} may only change \eif{sequence}, but not \eif{index} and not any component of the \eif{other} list's model.
\ifextended

The \eif{modify} clause mechanism is taken from other specification notations and methodologies (e.g., Spec\#~\cite{Specsharp}) usually targeted to formal correctness proofs.
It is only with a specification technique based on models, however, that it becomes practical for real classes and standard programming practices. 
Writing \eif{modify} clauses in terms of attributes would violate information hiding and be of limited usefulness to the client,
while listing arbitrary public queries is too tedious:
since the values of several regular queries are often related (for example, the value of \eif{off} may change when \eif{index} changes; see \cref{tab:llist_mbc}), \eif{modify} clauses should include all related queries, possibly also queries with arguments and on other objects.
Model queries are instead normally only a small number, they are orthogonal, and only depend on the state of the \eif{Current} object. 
Hence specifying which model queries change is not onerous; the values of all other queries are automatically defined in terms of them.
\fi

This approach to framing also supports a simple definition of specification completeness: a routine postcondition and framing specification are \emph{complete} if the relation between the model's pre- and poststate is a \emph{function}.\footnote{Such notion of completeness is of course relative to the model.}
Completeness is not an imperative in the MBC methodology:
programmers can still approach writing postconditions and framing incrementally.
It should rather be viewed as a safeguard against accidentally missing an important property.

\subsection{Class Invariants} \label{sec:class-invariants}

\noindent
\ifextended
  The \emph{class invariant} specifies global properties of valid instances of a class, which every operation must preserve.
  Since the semantics of class invariants can be subtle MBC introduces additional dedicated constructs for complex invariant properties.
  We borrow some ideas from the existing techniques developed for formal correctness proofs (e.g.,\xspace~\cite{Specsharp}, among many);
  unlike these sophisticated techniques, MBC's solution for class invariants does not target comprehensiveness, 
  but is easy to deploy and sufficient in practice for finding errors by testing and avoiding spurious invariant violations.

  \textbf{Class invariant types.}
  Like postconditions, class invariants in MBC use models to describe which object states are valid and which are not.
  For example, the first invariant clause in \cref{tab:llist_mbc} constrains the values of the model queries \eif{sequence} and \eif{index}, 
  stating that \eif{index} must never take values outside the interval $[0..\text{\eif{sequence.count} + 1}]$.

  Additionally, class invariants in MBC have three more specific usages: definitions of public attributes, linking invariants and internal representation constraints.
  Public attributes, from the class interface standpoint, are indistinguishable from public functions,
  and thus their values should be defined in terms of model queries.
  An example of such \emph{attribute definition} is the second invariant clause in \cref{tab:llist_mbc}, 
  which explains the attribute \eif{count} is terms of the model query \eif{sequence}.
   
  Parent classes may use simpler abstract models than their children. 
  \eif{LINKED_LIST}, for instance, inherits from a generic \eif{CONTAINER} class whose model is a bag (multiset) rather than a sequence, because the order of its elements is immaterial.
  To reuse the specification of the parent stated in terms of a different model, 
  we introduce class invariants that define the parent's model queries in terms of the child's model; 
  we call them \emph{linking invariants}.
  For example, the third invariant clause in \cref{tab:llist_mbc} says that the parent's model query \eif{bag} contains the same elements as \eif{sequence}, disregarding the order (\eif{sequence.to_bag}).

  Finally, \emph{internal representation constraints} introduce specifications that relate the values of model queries to the private attributes of the class.
  For example, the last invariant clause in \cref{tab:llist_mbc} says that the private attribute \eif{last_cell} stores the same value as \eif{sequence}'s last element (whenever the sequence is not empty).
  Unlike other MBC specifications,
  invariants of this type do not describe the public interface of the class 
  and usually cannot be made complete without revealing unnecessary implementation details in the model.
  However, even in this limited form, they turned out to be very effective at revealing errors that corrupt object's internal representation (see \cref{sec:faults-found}).
\fi

\textbf{Class invariant semantics.}
Eiffel checks class invariants at the beginning and at the end of every qualified\footnote{A call \eif{t.r} is \emph{qualified} when the target \eif{t} is an object other than \eif{Current}.} call on an object of the class.
This rule prevents checking the invariant whenever routines of a class call one another within the boundaries of a single object, in order to accomplish a common task,
as the object will normally be inconsistent (``open'') until all operations are completed.
\ifextended
  When circular dependencies between objects arise, this semantics may lead to spurious invariant violations: this is the \emph{dependent delegate} problem~\cite{ddd}.

  Consider an example derived from real code in \eb: 
  a binary tree data structure, where each node has a link to its \eif{parent} and \eif{left} and \eif{right} children.
  The \eif{Current} node is executing one of its routines and is temporarily in a state that violates the invariant; to restore it, it makes a qualified call on, say, its right child.
  The object \eif{right}, however, does not know that its \eif{parent} is in the middle of executing a call; if \eif{right} calls back to \eif{Current}, then, it detects an invariant violation even if \eif{right}'s call does not rely on the invariant.
  
  MBC deploys a runtime semantics where these spurious invariant violations do not occur.
  Objects are implicitly equipped with a Boolean attribute \eif{is_open} that is set to true at the entrance of every public routine call on the object and restored to its previous value when the routine terminates; class invariants are checked only if \eif{is_open} is false.
  This automatically solves the dependent delegate problem in the presence of callbacks: when \eif{right} calls back to \eif{Current}, the latter is open, and hence its invariant is not checked.

  This ``implicit opening'' mechanism is not sufficient to avoid spurious invariant violations when an object's invariant depends on the state of other objects.
  Consider again binary trees; an invariant states that the \eif{Current} node is its parent's left or right child:
  \begin{lstlisting}
    parent /= Void implies (parent.left = Current or parent.right = Current)
  \end{lstlisting}
  Routine \eif{prune_left} removes \eif{Current}'s left child as follows:
  \begin{lstlisting}
      old_left := left
      left := Void
      if old_left /= Void then old_left.set_parent (Void) end
  \end{lstlisting}
  When \eif{old_left.set_parent (Void)} is called to remove the back-link from \eif{Current}'s child, \eif{old_left}'s class invariant is violated: its parent's \eif{left} is already set to \eif{Void} and \eif{old_left} is not open; in fact, the very reason for calling \eif{set_parent} is to remove this inconsistency.
  MBC provides the keyword \eif{depend} to declare that an invariant clause depends on the state of an attribute, and hence it should be checked only if the object attached to attribute is closed.
  Annotating the invariant in the example with \eif{depend parent} removes the spurious invariant violation (\eif{old_left.parent} is \eif{Current}, which is open).
  
  In the few cases when fine-grained control over the opening of objects is necessary, MBC provides the \eif{open} clause for routines, 
  which \emph{explicitly} opens the objects attached to some of the routine's arguments when the routine begins execution 
  and restores them when the routine terminates 
  (as we discussed, the target is always opened implicitly).
  Consider a variant of the binary tree example where nodes have an attribute \eif{is_root} that should be true when their \eif{parent} node is \eif{Void}:
  \begin{lstlisting}
    parent = Void implies is_root = True
  \end{lstlisting}
  In this variant, \eif{prune} takes an argument of class \eif{NODE} that is supposed to be its left or right child and removes it as follows:
  \begin{lstlisting}
  prune (n: NODE)
    do
      if left = n then 
        left.set_parent (Void) ; left.set_root (True) ; left := Void 
      end
      if right = n then $\ldots$ end
    end
  \end{lstlisting}
  When \eif{prune}'s call to \eif{left.set_parent} returns, the invariant about \eif{parent} and \eif{is_root} is violated (\eif{left.parent = Void} but \eif{left.is_root} is still false). 
  Annotating \eif{prune} with \eif{open n} suspends checking of \eif{n}'s invariant until \eif{prune} terminates, thus removing the spurious invariant violation.  
\fi

As we discuss in \cref{sec:experiment_setup}, 
in \ebmbc we had to deploy explicit \eif{depend} and \eif{open} annotations only in a very few cases,
limited to doubly-linked list nodes, and binary and $n$-ary trees.

\subsection{Runtime Support for Strong Specifications} \label{sec:mbc_at_runtime}

\noindent
Model-based postconditions and invariants can be checked at runtime and used in testing out of the box:
with the same tools and user experience as standard Eiffel contracts.
Model queries introduced for specification purposes are implemented as regular functions
that compute the abstract model value from the concrete object state,
and thus do not require explicit initialization or updates. 
The specification classes we provide in MML are also regular Eiffel classes, implemented in a functional style.
Even though this approach to implementation of model queries and model classes potentially incurs a high runtime overhead,
the experiment results in \cref{sec:results} confirm that using MBC for contract-based testing is feasible.

Newly introduced specification constructs, such as \eif{modify}, \eif{depend} and \eif{open}, do not have any effect in the standard Eiffel semantics:
they are specified using \eif{note} meta-annotations (similar to Javadoc or C\#'s meta-data).
We have developed a simple tool that rewrites these annotations into plain Eiffel; 
for example, \eif{modify} clauses become explicit postconditions such as \eif{item = old item}. 
The MBC methodology is conservative, in that the class semantics is still sound if we ignore the special annotations; 
ignoring \eif{modify} clauses, for instance, yields weaker, yet correct, postconditions.

\section{Using Strong Specifications: Experiments} \label{sec:experiment_setup}

\noindent
We performed an extensive experimental evaluation to assess the benefits of using strong specifications for finding errors in software.

\subsection{Research Questions} \label{sec:research_questions}

\noindent
The overall goal of this evaluation is assessing and comparing the advantages and the cost of deploying strong specifications in the form of model-based contracts (MBC, described in \cref{sec:mbc}) 
when applied to automatic contract-based testing of real software.

This materializes into the following research questions:
\begin{enumerate}
\item Are strong specifications effective for finding faults in software?
\item Do strong specifications find subtle and complex faults?
\item Do strong specifications find faults in little testing time?
\item What is the performance overhead of checking strong specifications at runtime?
\item What is the development effort required to provide strong specifications for existing software?
\end{enumerate}

To answer these questions, we conducted two sets of experiments, targeting software written in Eiffel (\cref{sec:eiffel_setup}) and C\# (\cref{sec:csharp_setup}).
In both cases, we selected an open-source library, specified it following the MBC methodology, and extensively tested it with a standard automatic testing tool.
The rest of this section discusses the experiments; \cref{sec:results} presents the results.

\begin{table*}
\caption{Eiffel classes under test and results.}
\label{tab:class_overview}
\centering
\begin{threeparttable}
\begin{tabular}{l|rrrrrrr|rrrrrr}
\hline
\multicolumn{1}{c|}{\textsc{}} & \multicolumn{7}{c|}{\textsc{\eb}} & \multicolumn{6}{c}{\textsc{\ebmbc}}\\
\textsc{Class} & \textsc{LOC} & \textsc{PR} & \textsc{TC} & \textsc{Spec} & \textsc{Inc} & \textsc{Real} & \textsc{New} & \textsc{LOC} & \textsc{PR} & \textsc{TC} & \textsc{Inc} & \textsc{Real} & \textsc{New}\\
\hline
\rowcolor[gray]{0.9}\eif{ARRAY} & 831 & 53 & 2.8 & 2 & 0 & 2 & 1 & 986 & 59 & 1.2 & 0 & 3 & 2\\
\rowcolor[gray]{1.0}\eif{ARRAYED\_LIST} & 1840 & 86 & 3.5 & 0 & 0 & 0 & 0 & 2037 & 92 & 1.7 & 0 & 1 & 1\\
\rowcolor[gray]{0.9}\eif{ARRAYED\_QUEUE} & 537 & 32 & 1.8 & 0 & 0 & 2 & 0 & 648 & 37 & 3.8 & 0 & 2 & 0\\
\rowcolor[gray]{1.0}\eif{ARRAYED\_SET} & 1960 & 49 & 5.8 & 3 & 1 & 8 & 0 & 2053 & 58 & 5.4 & 0 & 16 & 8\\
\rowcolor[gray]{0.9}\eif{BINARY\_TREE} & 1122 & 64 & 1.0 & 2 & 5 & 6 & 0 & 1366 & 70 & 1.1 & 0 & 16 & 10\\
\rowcolor[gray]{1.0}\eif{BOUNDED\_QUEUE} & 558 & 32 & 1.4 & 0 & 0 & 2 & 0 & 659 & 37 & 3.8 & 0 & 2 & 0\\
\rowcolor[gray]{0.9}\eif{HASH\_TABLE} & 1345 & 51 & 0.9 & 1 & 0 & 1 & 0 & 1626 & 63 & 0.9 & 0 & 2 & 1\\
\rowcolor[gray]{1.0}\eif{HASH\_TABLE\_ITERATOR} & 217 & 15 & 0.4 & 0 & 0 & 0 & 0 & 248 & 15 & 0.5 & 0 & 0 & 0\\
\rowcolor[gray]{0.9}\eif{INDEXABLE\_ITERATOR} & 186 & 14 & 1.0 & 2 & 0 & 0 & 0 & 228 & 15 & 2.7 & 0 & 0 & 0\\
\rowcolor[gray]{1.0}\eif{INTEGER\_INTERVAL} & 519 & 42 & 4.3 & 1 & 1 & 0 & 0 & 637 & 45 & 0.9 & 0 & 3 & 3\\
\rowcolor[gray]{0.9}\eif{LINKED\_LIST} & 1759 & 69 & 2.0 & 0 & 0 & 2 & 0 & 1942 & 77 & 2.5 & 0 & 5 & 3\\
\rowcolor[gray]{1.0}\eif{LINKED\_LIST\_ITERATOR} & 311 & 15 & 0.7 & 0 & 0 & 0 & 0 & 357 & 16 & 0.7 & 0 & 0 & 0\\
\rowcolor[gray]{0.9}\eif{LINKED\_SET} & 2128 & 83 & 5.4 & 5 & 2 & 7 & 0 & 2410 & 94 & 4.8 & 0 & 24 & 17\\
\rowcolor[gray]{1.0}\eif{LINKED\_SET\_ITERATOR} & 311 & 15 & 0.7 & 0 & 0 & 0 & 0 & 357 & 16 & 0.7 & 0 & 0 & 0\\
\rowcolor[gray]{0.9}\eif{LINKED\_STACK} & 1077 & 27 & 1.0 & 0 & 0 & 3 & 1 & 1078 & 32 & 3.2 & 0 & 6 & 4\\
\rowcolor[gray]{1.0}\eif{TWO\_WAY\_LIST} & 2007 & 71 & 0.8 & 0 & 0 & 3 & 0 & 2184 & 79 & 2.2 & 0 & 6 & 3\\
\rowcolor[gray]{0.9}\eif{TWO\_WAY\_LIST\_ITERATOR} & 412 & 15 & 0.7 & 0 & 0 & 0 & 0 & 462 & 16 & 0.7 & 0 & 0 & 0\\
\rowcolor[gray]{1.0}\eif{TWO\_WAY\_SORTED\_SET} & 2706 & 91 & 5.3 & 5 & 2 & 9 & 0 & 2983 & 102 & 4.8 & 1 & 34 & 25\\
\rowcolor[gray]{0.9}\eif{TWO\_WAY\_SORTED\_SET\_ITERATOR} & 412 & 15 & 0.7 & 0 & 0 & 0 & 0 & 462 & 16 & 0.7 & 0 & 0 & 0\\
\rowcolor[gray]{1.0}\eif{TWO\_WAY\_TREE} & 2548 & 90 & 1.4 & 4 & 4 & 22 & 5 & 2865 & 101 & 1.3 & 0 & 29 & 12\\
\rowcolor[gray]{0.9}\eif{TWO\_WAY\_TREE\_ITERATOR} & 412 & 15 & 0.7 & 0 & 0 & 0 & 0 & 462 & 16 & 0.7 & 0 & 0 & 0\\
\hline
\rowcolor[gray]{1.0}\textbf{Total} & \textbf{17841} & \textbf{1033} & \textbf{42.5} & \textbf{15} & \textbf{12} & \textbf{48} & \textbf{7} & \textbf{19400} & \textbf{1164} & \textbf{44.4} & \textbf{1} & \textbf{103} & \textbf{62}\\
\hline
\end{tabular}
\begin{tablenotes}
\item \textsc{LOC}: Lines of code, \textsc{PR}: Public routines, \textsc{TC}: Test cases drawn (million)
\item \textsc{Spec}: Specification errors found, \textsc{Inc}: Inconsistency errors found, \textsc{Real}: Real faults found, \textsc{New}: Faults found only in this experiment
\end{tablenotes}
\end{threeparttable}

\end{table*}

\subsection{Eiffel Experiments} \label{sec:eiffel_setup}

\noindent
The main experiments target \eb (rev.~506)---Eiffel's standard base library---from which we selected 21 classes of varying size and complexity.
Using the facilities of the EiffelStudio IDE, we built the \emph{flat} version of each class, which is a self-contained implementation including all inherited members explicitly in the class text.
This simplified the task of writing specifications without being distracted by \eb's deep multiple inheritance hierarchy.
For each of the 21 classes in their flat version, \cref{tab:class_overview} lists the size (in \textsc{LOC}) and the number of public routines (\textsc{PR}), 
possibly also including helper classes directly used in the class implementation.
Since different classes may share some parent or helper classes, the totals at the bottom of the table are in general less than the sum of the elements in each column.

Like most Eiffel software, \eb comes with partial specification in the form of contracts: the 21 classes include 561 precondition clauses, 985 postcondition clauses, and 250 class invariant clauses.
In \ebmbc we completely replaced \eb's original postconditions and class invariants with model-based annotations,
but we kept \eb's preconditions (with a few exceptions discussed below)%
\footnote{All the code developed as part of the study, as well as descriptions of found faults are publicly available online~\cite{TSS-repository}.}.
\ifextended
\ebmbc's strong specification includes 589 precondition clauses, 1066 postcondition clauses and 164 class invariant clauses
(21\% model constraints, 23\% attribute definitions, 10\% linking invariants, 46\% internal representation constraints),
as well as 278 \eif{modify}, 4 \eif{depend} and 7 \eif{open} clauses.
\fi 
\cref{tab:class_overview} shows the size (in \textsc{LOC} and \textsc{PR}) of \ebmbc, 
which also includes model definitions and implementations of the model queries necessary to write MBC.

\textbf{Preconditions.}
In all but two \ebmbc classes we kept the same preconditions as in \eb.
Within the specific setup of our experiments, where we compare traditional contracts and strong contracts, 
it is important to have the same preconditions in the two artifacts under comparison.
Preconditions define the valid calling contexts of routines
(in particular, contract-based testing tools use them to select valid test cases).
Changing preconditions would change the semantics of classes in a way similar to changing implementation: 
strengthening a precondition may reduce the number of faults detectable for the routine, 
since it would move obligations from the routine to its clients; 
weakening a precondition may increase the number of faults,
since it would impose a heavier burden on its implementation.
We treat preconditions as developers' design decisions, which we normally take at face value. 
This policy makes the experiments with \eb and \ebmbc fully comparable.

The only exception occurred with four routines of class \eif{BINARY_TREE} and eight routines of class \eif{TWO_WAY_TREE} that insert new nodes into a tree.
In these twelve cases, we strengthened the preconditions to disallow creating cycles among nodes in the tree.
Without the strengthening, tree instances can be driven into inconsistent states with cycles where the whole specification of trees would be inapplicable. 
These changes in preconditions are conservative: 
the \ebmbc experiments using these stronger preconditions miss a few faults that are detected in \eb, 
because the new preconditions rule out some previously valid failing test cases.
Since these changes affect only a small fraction of all the experiments, the results with \eb and \ebmbc remain comparable.

\textbf{Specification correctness.}
To write correct strong contracts with MBC, we analyzed the original implementation, contracts, and comments in \eb, and relied on our informal knowledge of the semantics of data structures and their implementation.
To increase our confidence in the correctness of the new specification, we ran a series of short preliminary testing sessions with the goal of detecting inconsistencies and inaccuracies.
All our changes were conservative, in that whenever a new contract forbade a behavior that was not clearly forbidden by the comments, standard contracts, or informal knowledge, we weakened the specification to allow the behavior.
In all, we reached a high confidence that \ebmbc's specification is correct and strong enough.
The results of the main testing sessions (\cref{sec:results}) corroborate this informal assessment.

\textbf{Testing experiments.}
We ran a large number of random testing sessions with the AutoTest framework~\cite{bertrand2009} on a computing cluster of the Swiss National Supercomputing Centre, configured to allocate a standard 1.6~GHz core and 4~GB memory to each parallel AutoTest session.
The experiments totalled 1680 hours of testing time that generated nearly 87 millions of test cases; 
the \textsc{TC} columns in \cref{tab:class_overview} list the million of test cases drawn when testing each class in \eb and in \ebmbc.
The testing of every class was split into 30 sessions of 80 minutes, each with a new seed for the random number generator, such that corresponding sessions in \eb and \ebmbc use the same seeds.
This thorough testing protocol guaranteed statistically significant results~\cite{arcuri:practical:2011}.

\subsection{C\# Experiment} \label{sec:csharp_setup}

\noindent
A smaller set of experiments targets 9 classes from DSA (v.~0.6)---an open-source data structure and algorithm library written in C\#~\cite{DSA}.
Support for contracts in C\# appeared only recently, through the Code Contracts framework~\cite{CodeContracts}; 
therefore, most C\# projects (including DSA) do not have any formal specification.
This was a chance to extend the validation of the MBC methodology to other languages and to projects without pre-existing specification.

\begin{table}
\caption{C\# classes under test and results.}
\label{tab:csharp_class_overview}
\centering
\begin{threeparttable}
\begin{tabular}{l|rr|rr|rr}
\hline
\multicolumn{1}{c|}{\textsc{}} & \multicolumn{2}{c|}{\textsc{DSA}} & \multicolumn{2}{c|}{\textsc{DSA+}} & \multicolumn{2}{c}{\textsc{Testing}}\\
\textsc{Class} & \textsc{LOC} & \textsc{PR} & \textsc{LOC} & \textsc{PR} & \textsc{T} & \textsc{F}\\
\hline
\rowcolor[gray]{0.9}\csharp{AvlTree} & 345 & 6 & 391 & 7 & 23 & 1\\
\rowcolor[gray]{1.0}\csharp{BinarySearchTree} & 205 & 5 & 213 & 5 & 21 & 1\\
\rowcolor[gray]{0.9}\csharp{CommonBinaryTree} & 419 & 13 & 536 & 18 & 83 & 0\\
\rowcolor[gray]{1.0}\csharp{Deque} & 201 & 14 & 231 & 15 & 145 & 0\\
\rowcolor[gray]{0.9}\csharp{DoublyLinkedList} & 408 & 17 & 458 & 19 & 171 & 3\\
\rowcolor[gray]{1.0}\csharp{Heap} & 371 & 11 & 390 & 12 & 61 & 1\\
\rowcolor[gray]{0.9}\csharp{OrderedSet} & 136 & 9 & 158 & 11 & 10 & 0\\
\rowcolor[gray]{1.0}\csharp{PriorityQueue} & 186 & 13 & 216 & 14 & 65 & 0\\
\rowcolor[gray]{0.9}\csharp{SinglyLinkedList} & 439 & 20 & 492 & 22 & 148 & 3\\
\hline
\rowcolor[gray]{1.0}\textbf{Total} & \textbf{3043} & \textbf{133} & \textbf{3486} & \textbf{149} & \textbf{727} & \textbf{9}\\
\hline
\end{tabular}
\begin{tablenotes}
\item \textsc{LOC}: Lines of code, \textsc{PR}: Public routines
\item \textsc{T}: Testing time (minutes), \textsc{F}: Faults found
\end{tablenotes}
\end{threeparttable}

\end{table}

We instructed one of our bachelor's students to follow the methodology of \cref{sec:mbc} and create DSA+: a variant of DSA with the same implementation but equipped with strong model-based contracts.
DSA+'s specification includes 6 precondition clauses, 143 postcondition clauses and 23 class invariant clauses. 
For each of the 9 classes, \cref{tab:csharp_class_overview} shows the size (in \textsc{LOC} and \textsc{PR}) of both DSA and DSA+, inclusive of all specification elements and model query implementations.
As in \cref{tab:class_overview}, the count also includes (possibly shared) helper classes. 
Flattening was not necessary in this case because the inheritance hierarchy is shallow.

\textbf{Specification correctness.}
We manually inspected the DSA+ specification written by our student, and assessed its quality to be comparable to that of \ebmbc in terms of correctness and completeness.
Since DSA was not designed with contracts in mind, it makes recurrent usage of defensive programming, throwing exceptions to signal invalid arguments.
The experiment setup is consistent with this programming style: 
we do not consider such exceptions to be faults.

\textbf{Testing experiments.}
We performed automatic testing with the Pex concolic testing framework~\cite{TillmannH08} running on a Windows box equipped with a 2.16~GHz Intel Core2 processor and 3~GB of memory.
The experiments ran for about 12 hours; column \textsc{T} in \cref{tab:csharp_class_overview} reports the breakdown per class in minutes.
The testing time is different from class to class because Pex testing sessions by default are limited by coverage criteria rather than duration.
We only tested DSA+ since DSA has no formal specification elements usable as automated testing oracles.

\ifextended
  The C\# experiment is less extensive than the Eiffel experiment 
  and intended as a control mechanism to identify any potential dependency of the results on the Eiffel language, 
  libraries (\eb) or tools.
\fi

\section{Using Strong Specifications: Results} \label{sec:results}

\noindent
This section discusses the result of the experiments, focusing on the larger \eb experiments, 
with \ref{sec:faults-found} through \ref{sec:spec-writ-overh} targeting the research questions 1--5 of \cref{sec:research_questions}. 
Then, \ref{sec:c-experiment} briefly discusses the experiments with C\#,
and \ref{sec:threats-validity} presents possible threats to validity of the results.

\subsection{Faults Found} \label{sec:faults-found}

\noindent
AutoTest found 75 faults in \eb and 104 in \ebmbc; these are \emph{unique}, that is they identify distinct and independent errors.
We classified them in three categories.

\emph{Specification faults} correspond to violations of \emph{wrong} contracts 
(meaning that in our judgement they specify the expected behavior of the program incorrectly).
We found 15 specification faults in \eb (column \textsc{Spec} in \cref{tab:class_overview}) and none in \ebmbc, 
which increased our confidence that the preliminary testing sessions mentioned in \cref{sec:eiffel_setup} were sufficient to achieve correct specifications.
We consider specification faults spurious in our study, because we are not comparing the correctness of the specification in \eb and \ebmbc but rather their effectiveness at finding real errors in the implementation.

\emph{Inconsistency faults} correspond to failures triggered by calls on objects in inconsistent states,
which are not captured by a partial class invariant.
For example, \eif{LINKED_SET} may be driven into a state where the container stores duplicate elements; calling \eif{remove (x)} in such a state triggers a failure (only one occurrence of \eif{x} is removed), but \eif{remove} is not to blame for it, since it is due to previous erroneous behavior that went undetected.
While inconsistency faults are genuine errors, we classify them separately because understanding and locating the ultimate source of an inconsistency is normally harder.
Additionally, a single inconsistency fault often results in many failing test cases (potentially in all routines of the class that rely on the broken invariant),
requiring additional effort from the developer when analyzing the testing results.

We found 12 inconsistency faults in \eb and 1 in \ebmbc (columns \textsc{Inc} in \cref{tab:class_overview}); 
the ultimate source of the latter fault is a class invariant not including all internal representation constraints (see \cref{sec:class-invariants}), 
which would have required exposing implementation details in the model.
The other inconsistency faults of \eb are not detected in \ebmbc,
because, due to stronger class invariants, their \emph{real} source is detected instead.
In the \eif{LINKED_SET} example above, instead of the inconsistency fault in \eif{remove},
MBC report a fault in routine \eif{replace}, which does not check if the new value is already present in the set, thereby introducing duplicates. 
The results in this category indicate that strong specifications report faults in a way that is easier to understand and debug.

All other errors are \emph{real faults} which correspond to genuine errors directly traceable to the code.
We found 48 real faults in \eb and 103 in \ebmbc (columns \textsc{Real} in \cref{tab:class_overview}); 41 of them are found in both sets of experiments, 7 only in \eb, and 62 only in \ebmbc.
We submitted bug reports for all the 110 faults found in our experiments.
The Eiffel Software developers in charge confirmed 107 (97\%) of them as real bugs to be fixed.
This is evidence that we are dealing with genuine faults in our evaluation.
The remaining three faults not taken on by the developers also arguably highlight real problems in the implementation, but they are probably not so likely to occur during ``normal'' runs.
The rest of the discussion focuses on real faults unless stated otherwise.

Only seven faults are found in \eb but not in \ebmbc (columns \textsc{New} in \cref{tab:class_overview}).
Four of them are prevented by the strengthened preconditions in the tree classes (\cref{sec:eiffel_setup}); 
two are shadowed by new failures occurring earlier; 
and one disappears with MBC due to an unintentional side-effect of a model query that amends an invariant violation.
None of these faults found only in \eb show inherent deficiencies of strong specifications or of the MBC method.
In contrast, the 62 faults found only in \ebmbc are undetectable in \eb.

\begin{figure}
\centering
\includegraphics[width=\columnwidth]{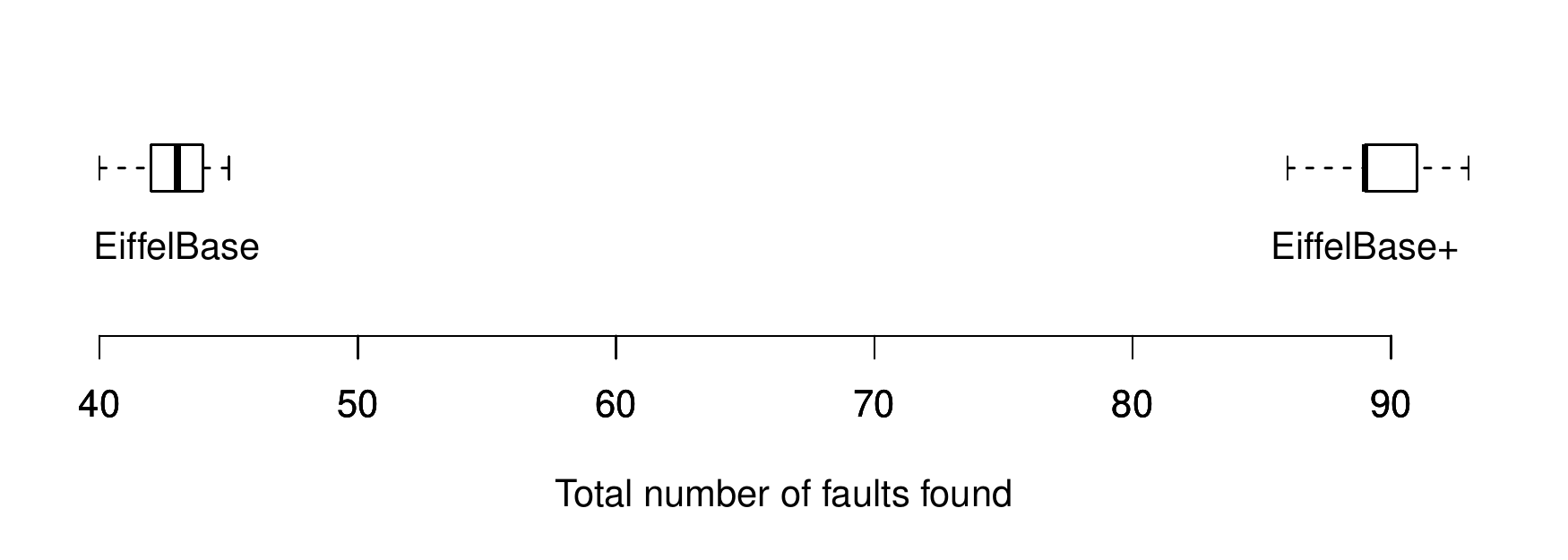}
\caption{Unique real faults found in all classes over 80-minute testing sessions.} \label{fig:faults_boxplot}
\end{figure}

Except for the two \eif{ITERATOR} classes (no faults in both cases) and the two \eif{QUEUE} classes (the same two faults in both cases), the number of faults found is consistently higher in \ebmbc \emph{in each class}.
As evident from the boxplot in \cref{fig:faults_boxplot}, the difference is highly significant:
the Mann-Whitney $U$ test gives $U = 0$ (testing \ebmbc outperforms testing \eb in \emph{all} sessions), 
and $p = 2 \cdot 10^{-11}$ overall and $p \leq 2.1 \cdot 10^{-11}$ for every class (except the \eif{ITERATOR}s and \eif{QUEUE}s).
The difference remains highly statistically significant even if we aggregate the experiments in sessions of different length.

\begin{center}
\framebox{\parbox{8.2cm}{\centering Testing with strong specifications detected 55 more (twice as many) unique real faults than testing with standard, partial contracts. 62 (56\%) of the faults are detected \emph{only} with strong specifications.}}
\end{center}

\subsection{Fault Complexity} \label{sec:fault-complexity}

\noindent
Although it is to some extent subjective whether a fault is ``deep'' or ``subtle'', 
faults violating postconditions or class invariants are arguably more complex 
because so are the violated properties.
While there is no significant difference in the percentage of class invariant violations between \eb and \ebmbc (33\% in both cases), 
postconditions trigger 42\% of violations in \ebmbc but only 11\% in \eb: 
the Wilcoxon signed-rank test among all classes gives $W = 0$ and $p = 6 \cdot 10^{-3}$ both for postconditions alone and for postconditions and class invariants counted together, 
which demonstrates that strong specifications systematically detect more complex errors.
76\% of faults in \ebmbc are detected thanks to postconditions or invariants---a direct consequence of the effectiveness of the MBC methodology for writing them.

\ifextended
  One example of a fault detected by a model-based postcondition was already discussed in \cref{sec:examples}.
  Here we give two other examples to demonstrate that they are indeed subtle yet understandable:
  \begin{itemize}
  \item Routine \eif{ARRAY.force(v, i)} inserts value \eif{v} at position \eif{i} into an array, extending its bounds if needed.
  All elements in between the old bound and \eif{i} are supposed to be initialized with default values,
  however \eif{force} contains an off-by-one error,
  and in a particular scenario fails to initialize one element.
  This is missed by the original postcondition \eif{item(i) = v}, which only takes care of the newly inserted element,
  but detected by the complete model-based postcondition, 
  which, following the methodology, specifies array elements at all positions.
  \item Both \eif{ARRAYED_SET} and \eif{LINKED_SET} inherit most of their implementation from the corresponding list classes,
  including the implementation of \eif{is_equal}: the object equality function.
  As a result, two sets with the same elements in a different order are considered different.
  The original postcondition only states that equal sets must have the same size and that equality is symmetric,
  which does not capture the specifics of set equality.
  \end{itemize}

  It is revealing that 11 faults in \ebmbc are detected due to violations of contracts generated automatically by our tool that processes MBC annotations (\cref{sec:mbc_at_runtime}) such as \eif{modify} and \eif{depend}.
\fi
These faults are practically out of the scope of regular contracts,
as specifying the corresponding properties explicitly is extremely onerous.

\ifextended
  Throughout the whole experiment we encountered one violation of an invariant
  that could be later restored before the enclosing public routine call terminates.
  Strictly speaking, such violation is spurious,
  and to eliminate it we would have to extend the notation for \eif{open} clauses,
  in order to support opening arbitrary expressions rather than just routine arguments.
  However in reality this particular invariant was \emph{not} restored,
  so the violation pointed to a real fault.
  This example suggests that if an object is too ``far away'' in the object structure from the call target
  to be mentioned in the \eif{open} or \eif{depend} clause,
  it is likely that a developer forgets to restore its invariant anyway,
  because the object is not in the area of immediate interest for the routine. 
\fi

\begin{figure}
\centering
\includegraphics[width=\columnwidth]{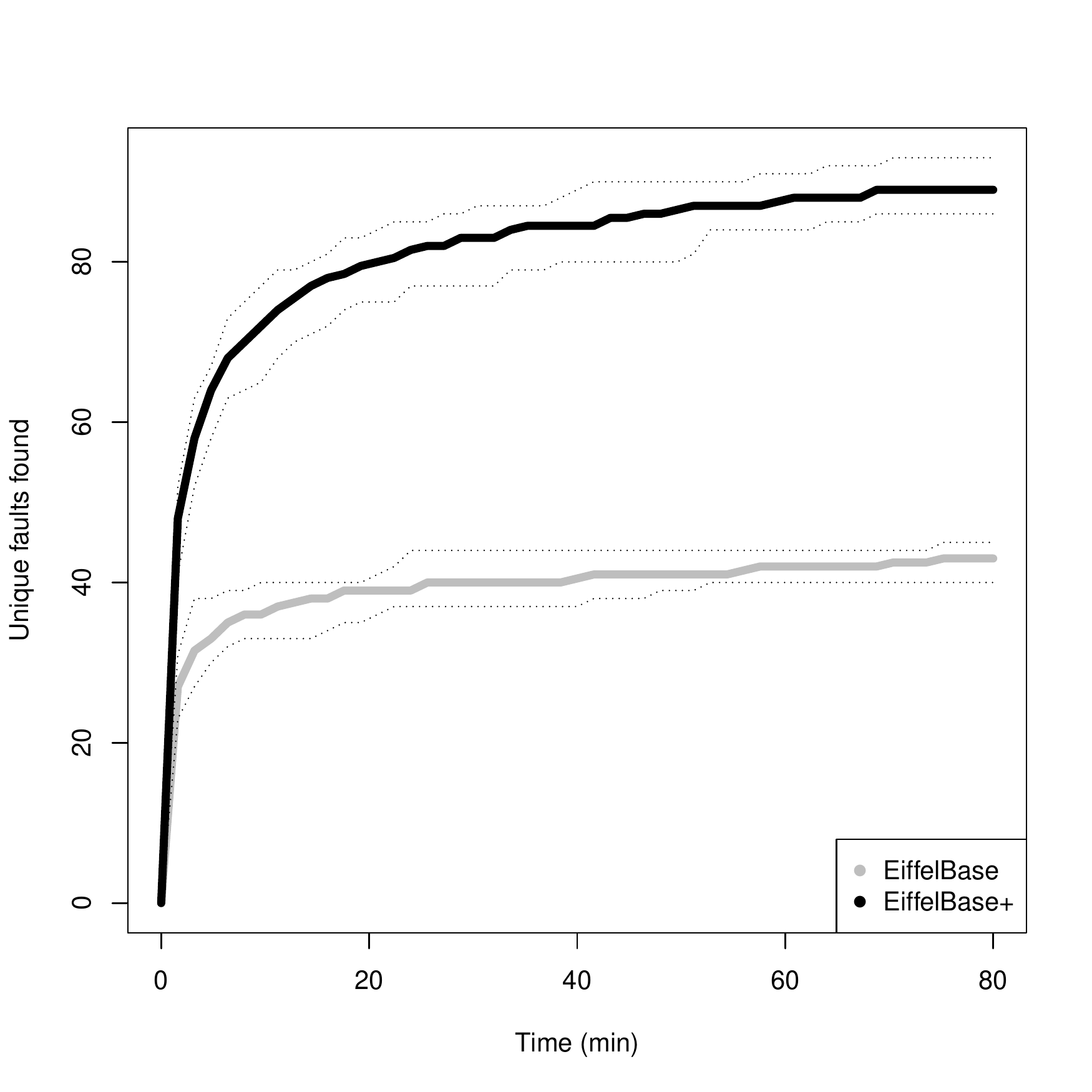}
\caption{Median number of faults, aggregated from all classes, in time. Dotted lines show minimum and maximum for each case.} \label{fig:faults_in_time}
\end{figure}

\subsection{Usage of Testing Time} \label{sec:usage-testing-time}

\noindent
\cref{fig:faults_in_time} plots the number of faults detected in \eb and \ebmbc over a median 80-minute session; it is clear that the behavior with strong specifications dominates over standard contracts after only a few minutes.
Dominance is observed consistently in all classes (with the usual exception of \eif{ITERATOR}s and \eif{QUEUE}s): a median session with strong contracts finds more faults than a median session with standard contracts after a time between two seconds and five minutes
depending on the class under test;
after a time between 13 seconds and 20 minutes, testing with strong contracts finds more faults than testing with standard contracts will find in the whole session.

Testing with standard contracts also seems to exhaust earlier its fault-finding potential: given any time from 20 minutes on, there are more \eb sessions than \ebmbc sessions that have found all the faults they ever will by this time.
This may indicate that standard contracts are good to find ``quick to detect'' faults, but they also soon run out of steam.

We considered other differences between experiments with \eb and with \ebmbc in the usage of testing time: repeatability of testing session history, and the presence of \emph{rare} faults triggered only in a small number of cases.
Our experiments with strong specifications are slightly less repeatable and include a few more rare faults, but the differences with standard contracts are not statistically significant.

\subsection{Runtime Performance Overhead} \label{sec:runt-perf-overh}

\noindent
Runtime checking of strong specifications based on models often requires traversing the whole data structure to construct an object of a model class, whenever a contract element is exercised.
As a rule, this demands more computational resources than executing the simple checks involved in standard contracts.
To measure the runtime overhead of checking MBC specifications in automated testing, we compared the number of test cases generated by AutoTest in the same amount of time when testing \eb and \ebmbc.
Contrary to our expectations, the overhead is small in many cases and not significant overall (see column \textsc{TC} of \cref{tab:class_overview}). 
A possible interpretation of this data is that the overhead of strong specifications grows as larger data structures are instantiated; 
because random testing most of the time only exercises small data structures, this overhead does not show.

We did not find a significant correlation between the variation of overhead for different classes and any source code metrics we considered. 
On the other hand, some AutoTest heuristics that decide to discard previously created objects are activated more often for classes where strong specifications are faster to check.

\begin{table}
\centering
\caption{Specification overhead} \label{tab:overhead}
\begin{tabular}{l|rrr}
\hline
\textsc{\# tokens} & \textsc{\eb} & \textsc{\ebmbc} & \textsc{Overhead}\\
\hline
\rowcolor[gray]{0.9}Preconditions & 1514 & 1696 & 1.12\\
\rowcolor[gray]{1.0}Postconditions & 5410 & 11837 & 2.19\\
\rowcolor[gray]{0.9}Invariants & 1508 & 1587 & 1.05\\
\rowcolor[gray]{1.0}MBC annotations &  & 1893 & \\
\rowcolor[gray]{0.9}Model queries &  & 2268 & \\
\hline
\rowcolor[gray]{1.0}\textbf{Total} & \textbf{8432} & \textbf{19281} & \textbf{2.29}\\
\rowcolor[gray]{0.9}Spec/code & 0.20 & 0.46 & \\
\hline
\end{tabular}

\end{table}

\subsection{Specification Writing Overhead} \label{sec:spec-writ-overh}

\noindent
Applying MBC to create \ebmbc required roughly one person-month, plus one person-week of preliminary testing for fine-tuning the specification,
which puts the overall ratio benefit/effort at about four defects detected per person-day. 
\cref{tab:overhead} measures the amount of work produced in this time:
for each specification item, including preconditions, postconditions, class invariants, MBC annotations such as \eif{modify}, and model query implementations, we compare the number of \emph{tokens} in \ebmbc against those in \eb (when applicable) and give the \textsc{Overhead} of strong specifications as the ratio of the two values.
The last line also shows the overall specification to code ratios.

Reflecting the importance MBC gives to strong postconditions and the more restricted role of class invariants, 67\% of all \emph{new} specification in \ebmbc are postconditions, whereas only 9\% are class invariants.
MBC-specific annotations are 11\%, mostly \eif{modify} clauses that are however straightforward to write and dispense for more intricate explicit framing specifications.
Model query implementations account for the remaining 13\%.

These numbers suggest that the specification overhead of MBC is moderate and abundantly paid off by the advantages in terms of errors found and quality of documentation.
The specification to code ratio also compares favorably to other approaches to improving software quality.
Detailed quantitative data about industrial experiences with test-driven development is scarce, but few references indicate~\cite{TDD-book,TDD-empirical,MW-ICSE03} that it is common to have between 0.4 and 1.0 lines of tests per line of application code for projects of size comparable to \eb.
Correctness proofs are normally much more demanding, as they require between 1.5 and 9 specification elements per implementation element~\cite{VSComp2010, Dupressoir11, Filliatre12}.

\subsection{C\# Experiments} \label{sec:c-experiment}

\noindent
Pex found 9 unique faults in DSA+ violating the model-based specification (column \textsc{F} in \cref{tab:csharp_class_overview}).
Unfortunately, we could not get an evaluation of these faults by the original code developers.
We have confidence, however, that the faults uncover some obvious errors and, even in the most benign interpretation, some instances of bad object-oriented design.

The fault rates (faults per line of executable code) are comparable in the Eiffel and C\# experiments, being respectively $6\cdot 10^{-3}$ and $3\cdot 10^{-3}$.
The fault complexity is also qualitatively similar for the two languages.
The testing time (column \textsc{T} in \cref{tab:csharp_class_overview}) is instead incomparable, as Pex and AutoTest implement very different testing algorithms.

Applying MBC to create DSA+ required roughly 50 person-hours, plus another 8 person-hours used by the student to learn the MBC methodology on small examples.
The specification/code ratio is perceptibly higher in DSA+ compared to \ebmbc (0.9); this is largely due to the verbose syntax of Code Contracts which are a library, as opposed to Eiffel's native language support for contracts.

\subsection{Threats to Validity} \label{sec:threats-validity}

\noindent
Threats to internal validity of our findings come from the usage of randomized testing tools, whose behavior may change in different sessions.
We designed the experimental protocol~\cite{arcuri:practical:2011} to reduce this threat to a minimum: we ran a large number of repeated experiments and we performed suitable non-parametric statistical tests of significance for all differences we observed.

Threats to external validity refer to the generalizability of our findings.
While MBC leads to very good results in our experiments, applying it to programs in application domain other than data structures might be more difficult or require an extension of the technique.
Our results remain significant, however, if compared to the state of the art in deploying strong specifications.
The generalizability to other languages and analysis tools is partially addressed by our experiments targeting two languages (Eiffel and C\#) and two automatic testing technologies (random and concolic).
Future work will experiment with even more approaches and notations.

\section{Related Work} \label{sec:related_work}

\noindent
This section discusses the most significant related work in three areas: using formal specifications for testing; using inferred specifications to improve testing; and model-based specification methods.

\textbf{Formal specifications for testing.}
The idea of using formal specifications for testing has a history that stretches back more than three decades; see~\cite{HieronsBBCDDGHKKLSVWZ09} for a comprehensive survey.
Various proposals targeted different specification formalisms including algebraic datatypes~\cite{GannonMH81,ChangR99}, logic-based notations~\cite{StocksC93}, UML Statecharts~\cite{OffuttA99} and other state machines, and contracts and similar forms of embedded assertions~\cite{MarinovK01,CheonL02,BoyapatiKM02,bertrand2009}.
In these applications, formal specifications provide reliable---often automated---testing oracles~\cite{StaatsWH11} and can also guide test planning and test case generation.

This extensive experience is evidence that formal specifications can improve the testing process.
From a software engineering viewpoint, however, an outstanding open issue is finding optimal trade-offs between the effort required to provide formal specifications and the improvements (in efficiency and effectiveness) they bring to the testing of real software.
The evidence---empirical~\cite{MullerTH02} or anecdotal~\cite{Parnas-FOSE}---is scarce in this area: 
most successful experiences do not explicitly take into account the effort required to produce reliable specifications against the benefits gained for testing (e.g., \cite{BousquetLMOL10}); 
or they only target partial specifications, 
which have the advantage of being easy to write (e.g., \cite{BoyapatiKM02,bertrand2009}).
In contrast, this paper targeted the high-hanging fruit of deploying strong specifications, 
explicitly addressing
the difficulties of writing and using such specifications for existing software.
Our results that strong specifications reveal complex (design) errors corroborate Hoare's view that the real value of tests is that ``they detect inadequacy in the [development] methods''~\cite{Hoare96}.

\textbf{Inferred specifications for testing.}
When specifications can be inferred automatically from the code, the deployment effort is negligible compared to the benefits they bring.
Therefore, a number of recent works (e.g., \cite{PachecoE05,XieN06,Zeller10,WRFPHSNM-ASE11}) developed sophisticated techniques for inferring specifications from program executions with the intent of using them to improve testing.
The experiments reported in these papers show that inferred specifications can boost automated testing~\cite{dAmorimPXME06}; on the other hand, even the most accurate inferred specifications only express the code from a different angle, and hence cannot take the developer's intent fully into account and are necessarily limited to detecting certain types of inconsistencies.
Combining inferred and manually written specifications is an interesting endeavor that belongs to future work (see~\cite{Polikarpova2009,WFKM11-ICSE11} for some preliminary studies).

\textbf{Model-based specification methods.}
The methodology described in \cref{sec:spec_methodology} extends our previous work~\cite{PFM10-VSTTE10} with the specific goal of developing executable specifications for automated testing.
The same goal has also motivated the techniques to improve the runtime checking of strong specifications described in \cref{sec:mbc_at_runtime}.
The related work section of~\cite{PFM10-VSTTE10} compares the foundations of our model-based method against other similar approaches such as JML~\cite{Leavens2005}.

\section{Conclusions and Future Work} \label{sec:conclusions}

\noindent
This paper presents a methodology to write strong specifications that extends the traditional Design by Contract, and applied it to specifying data structure classes in Eiffel and C\#.
We carried out an extensive empirical evaluation to determine the benefits of using such strong specifications in testing with automatic tools.
We found twice as many bugs in the software with strong specifications as in the same software specified with standard partial contracts.
We also demonstrated that the effort required to write the strong specifications was moderate thanks to the methodology that is practical and palatable to professionals not fluent in formal techniques.

The benefits brought by strong specifications are not limited to finding errors through testing.
While the present paper focused on adding strong specifications to existing code a posteriori, our related work~\cite{PFM10-VSTTE10} shows that model-based contracts help achieve consistent designs and higher-quality code by construction.

As \textbf{future work}, we plan to extend the MBC methodology and supporting tools to work on more complicated application domains with a higher degree of automation, and to support other software analysis techniques such as correctness proofs and static analysis.
We will also expand the experimental evaluation to more projects and programming languages. 

{\vskip 2pt}
\textbf{Acknowledgements.} Thanks to Rosemary Monahan and Scott West for comments on a draft of this paper; and to Tobias Kiefer for carrying out the C\# experiments.
This work was partially supported by the Swiss SNF under projects FullContracts (200021-137931) and ASII (200021-134976); we also acknowledge the support of the Swiss National Supercomputing Centre for the testing experiments.

\bibliographystyle{IEEEtran}
\bibliography{misc_bibliography,mbc_bibliography,mml_relatedwork}

\begin{thebibliography}{10}
\providecommand{\url}[1]{#1}
\csname url@samestyle\endcsname
\providecommand{\newblock}{\relax}
\providecommand{\bibinfo}[2]{#2}
\providecommand{\BIBentrySTDinterwordspacing}{\spaceskip=0pt\relax}
\providecommand{\BIBentryALTinterwordstretchfactor}{4}
\providecommand{\BIBentryALTinterwordspacing}{\spaceskip=\fontdimen2\font plus
\BIBentryALTinterwordstretchfactor\fontdimen3\font minus
  \fontdimen4\font\relax}
\providecommand{\BIBforeignlanguage}[2]{{%
\expandafter\ifx\csname l@#1\endcsname\relax
\typeout{** WARNING: IEEEtran.bst: No hyphenation pattern has been}%
\typeout{** loaded for the language `#1'. Using the pattern for}%
\typeout{** the default language instead.}%
\else
\language=\csname l@#1\endcsname
\fi
#2}}
\providecommand{\BIBdecl}{\relax}
\BIBdecl

\bibitem{Parnas-FOSE}
D.~L. Parnas, ``Precise documentation: The key to better software,'' in
  \emph{The Future of Software Engineering}.\hskip 1em plus 0.5em minus
  0.4em\relax Springer, 2011, pp. 125--148.

\bibitem{Meyer97}
B.~Meyer, \emph{Object Oriented Software Construction}.\hskip 1em plus 0.5em
  minus 0.4em\relax Prentice Hall, 1997.

\bibitem{Chalin06}
P.~Chalin, ``Are practitioners writing contracts?'' in \emph{The RODIN Book},
  ser. LNCS, vol. 4157, 2006, p. 100.

\bibitem{TDD-book}
K.~Beck, \emph{Test-Driven Development}.\hskip 1em plus 0.5em minus 0.4em\relax
  Addison-Wesley, 2002.

\bibitem{PFM10-VSTTE10}
N.~Polikarpova, C.~A. Furia, and B.~Meyer, ``Specifying reusable components,''
  in \emph{VSTTE}, ser. LNCS, vol. 6217, 2010, pp. 127--141.

\bibitem{bertrand2009}
B.~Meyer, A.~Fiva, I.~Ciupa, A.~Leitner, Y.~Wei, and E.~Stapf, ``Programs that
  test themselves,'' \emph{IEEE Computer}, vol.~42, no.~9, pp. 46--55, 2009.

\bibitem{DSA}
http://dsa.codeplex.com/.

\bibitem{TillmannH08}
N.~Tillmann and J.~de~Halleux, ``Pex--white box test generation for {.NET},''
  in \emph{TAP}, 2008, pp. 134--153.

\bibitem{Polikarpova2009}
N.~Polikarpova, I.~Ciupa, and B.~Meyer, ``A comparative study of
  programmer-written and automatically inferred contracts,'' in \emph{ISSTA},
  2009, pp. 93--104.

\bibitem{Specsharp}
M.~Barnett, K.~R.~M. Leino, and W.~Schulte, ``The spec\# programming system: an
  overview,'' in \emph{CASSIS}.\hskip 1em plus 0.5em minus 0.4em\relax Berlin,
  Heidelberg: Springer-Verlag, 2005, pp. 49--69.

\bibitem{ddd}
B.~Meyer, ``The dependent delegate dilemma,'' in \emph{Engineering Theories of
  Software Intensive Systems}.\hskip 1em plus 0.5em minus 0.4em\relax Springer,
  2005.

\bibitem{TSS-repository}
http://se.inf.ethz.ch/people/polikarpova/mbctesting.

\bibitem{arcuri:practical:2011}
A.~Arcuri and L.~Briand, ``A practical guide for using statistical tests to
  assess randomized algorithms in software engineering,'' in \emph{ICSE}.\hskip
  1em plus 0.5em minus 0.4em\relax ACM, 2011, pp. 1--10.

\bibitem{CodeContracts}
http://research.microsoft.com/en-us/projects/contracts/.

\bibitem{TDD-empirical}
N.~Nagappan, E.~M. Maximilien, T.~Bhat, and L.~Williams, ``Realizing quality
  improvement through test driven development: results and experiences of four
  industrial teams,'' \emph{ESE}, vol.~13, no.~3, pp. 289--302, 2008.

\bibitem{MW-ICSE03}
E.~M. Maximilien and L.~Williams, ``Assessing test-driven development at
  {IBM},'' in \emph{ICSE}, 2003, pp. 564--569.

\bibitem{VSComp2010}
V.~Klebanov \emph{et~al.}, ``The 1st verified software competition,'' in
  \emph{FM}, ser. LNCS, vol. 6664, 2011, extended version at
  \url{www.vscomp.org}.

\bibitem{Dupressoir11}
F.~Dupressoir, A.~D. Gordon, J.~J\"urjens, and D.~A. Naumann, ``Guiding a
  general-purpose {C} verifier to prove cryptographic protocols,'' in
  \emph{IEEE Computer Security Foundations Symposium}, 2011.

\bibitem{Filliatre12}
J.-C. Filli\^{a}tre, ``Verifying two lines of {C} with {Why3}: an exercise in
  program verification,'' in \emph{VSTTE}, ser. LNCS, 2012, pp. 83--97.

\bibitem{HieronsBBCDDGHKKLSVWZ09}
R.~M. Hierons \emph{et~al.}, ``Using formal specifications to support
  testing,'' \emph{ACM Comput. Surv.}, vol.~41, no.~2, 2009.

\bibitem{GannonMH81}
J.~D. Gannon, P.~R. McMullin, and R.~G. Hamlet, ``Data-abstraction
  implementation, specification, and testing,'' \emph{ACM Trans. Program. Lang.
  Syst.}, vol.~3, no.~3, pp. 211--223, 1981.

\bibitem{ChangR99}
J.~Chang and D.~J. Richardson, ``Structural specification-based testing:
  Automated support and experimental evaluation,'' in \emph{ESEC/FSE}, 1999,
  pp. 285--302.

\bibitem{StocksC93}
P.~Stocks and D.~A. Carrington, ``Test templates: A specification-based testing
  framework,'' in \emph{ICSE}, 1993, pp. 405--414.

\bibitem{OffuttA99}
A.~J. Offutt and A.~Abdurazik, ``Generating tests from {UML} specifications,''
  in \emph{UML}, 1999, pp. 416--429.

\bibitem{MarinovK01}
D.~Marinov and S.~Khurshid, ``{TestEra}: A novel framework for automated
  testing of {J}ava programs,'' in \emph{ASE}, 2001, p.~22.

\bibitem{CheonL02}
Y.~Cheon and G.~T. Leavens, ``A simple and practical approach to unit testing:
  The {JML} and {JUnit} way,'' in \emph{ECOOP}, 2002, pp. 231--255.

\bibitem{BoyapatiKM02}
C.~Boyapati, S.~Khurshid, and D.~Marinov, ``{K}orat: automated testing based on
  {J}ava predicates,'' in \emph{ISSTA}, 2002, pp. 123--133.

\bibitem{StaatsWH11}
M.~Staats, M.~W. Whalen, and M.~P.~E. Heimdahl, ``Programs, tests, and
  oracles,'' in \emph{ICSE}, 2011, pp. 391--400.

\bibitem{MullerTH02}
M.~M. M{\"u}ller, R.~Typke, and O.~Hagner, ``Two controlled experiments
  concerning the usefulness of assertions as a means for programming,'' in
  \emph{ICSM}, 2002, pp. 84--92.

\bibitem{BousquetLMOL10}
L.~du~Bousquet, Y.~Ledru, O.~Maury, C.~Oriat, and J.-L. Lanet, ``Reusing a
  {JML} specification dedicated to verification for testing, and vice-versa:
  Case studies,'' \emph{J. Autom. Reasoning}, vol.~45, no.~4, pp. 415--435,
  2010.

\bibitem{Hoare96}
C.~A.~R. Hoare, ``How did software get so reliable without proof?'' in
  \emph{FME}, 1996, pp. 1--17.

\bibitem{PachecoE05}
C.~Pacheco and M.~D. Ernst, ``{Eclat}: Automatic generation and classification
  of test inputs,'' in \emph{ECOOP}, 2005, pp. 504--527.

\bibitem{XieN06}
T.~Xie and D.~Notkin, ``Tool-assisted unit-test generation and selection based
  on operational abstractions,'' \emph{Autom. Softw. Eng.}, vol.~13, no.~3, pp.
  345--371, 2006.

\bibitem{Zeller10}
A.~Zeller, ``Mining specifications: A roadmap,'' in \emph{The Future of
  Software Engineering}.\hskip 1em plus 0.5em minus 0.4em\relax Springer, 2010,
  pp. 173--182.

\bibitem{WRFPHSNM-ASE11}
Y.~Wei, H.~Roth, C.~A. Furia, Y.~Pei, A.~Horton, M.~Steindorfer, M.~Nordio, and
  B.~Meyer, ``Stateful testing: Finding more errors in code and contracts,'' in
  \emph{ASE}, 2011, pp. 440--443.

\bibitem{dAmorimPXME06}
M.~d'Amorim, C.~Pacheco, T.~Xie, D.~Marinov, and M.~D. Ernst, ``An empirical
  comparison of automated generation and classification techniques for
  object-oriented unit testing,'' in \emph{ASE}, 2006, pp. 59--68.

\bibitem{WFKM11-ICSE11}
Y.~Wei, C.~A. Furia, N.~Kazmin, and B.~Meyer, ``Inferring better contracts,''
  in \emph{ICSE}, 2011, pp. 191--200.

\bibitem{Leavens2005}
G.~T. Leavens, Y.~Cheon, C.~Clifton, C.~Ruby, and D.~R. Cok, ``How the design
  of {JML} accommodates both runtime assertion checking and formal
  verification,'' \emph{Sci. Com. Prg.}, vol.~55, no. 1-3, pp. 185--208, 2005.

\end{thebibliography}

\end{document}